\documentclass[aps,prl,twocolumn,groupedaddress]{revtex4}
\usepackage{amsmath}
\usepackage{amssymb}
\usepackage{graphicx}

\begin{document}

\title{Pseudospin in optical and transport properties of graphene}

\author{Maxim Trushin and John Schliemann}

\affiliation{Institute for Theoretical Physics, University of Regensburg,
D-93040 Regensburg, Germany}

\date{\today}

\begin{abstract}
We show that the pseudospin being an additional degree of freedom
for carriers in graphene can be efficiently controlled
by means of the electron-electron interactions which, in turn,
can be manipulated by changing the substrate.
In particular, an out-of-plane pseudospin component can occur
leading to a zero-field Hall current as well as to 
polarization-sensitive interband optical absorption.
\end{abstract}

\maketitle

{\em Introduction.} 
The charge carriers in graphene  are described at low energies
by an effective Hamiltonian being formally equivalent to the massless 
two-dimensional Dirac Hamiltonian\cite{Nature2005novoselov,Nature2007geim,RMP2009castroneto},
$H^{\nu}_0=\hbar v_0 (\nu\sigma_x k_x+\sigma_y k_y)$,
where $\nu=\pm$ refers to the two inequivalent corners $K$, $K'$
of the first Brillouin zone, $v_0\approx 10^6 \mathrm{ms^{-1}}$ is the effective ``speed of light'',
${\mathbf k}$ is the two-component particle momentum operator, and
$\sigma_{x,y}$ are Pauli matrices describing the sublattice degree of freedom
also referred to as the pseudospin\cite{RMP2009castroneto}.
In the original Dirac Hamiltonian the Pauli vector $\vec{\sigma}$
represents the spin of a spin-$1/2$ particle which can
be detected in Stern--Gerlach-like experiments.
The pseudospin in graphene is formally similar to the true
electron spin with an important distinction given by the
behavior under time and parity \cite{PRL2005kane} inversion.
For the above effective model the
time reversal operator $T$ is just the operator $C$ of complex conjugation,
$T=C$, fulfilling $TH^{\nu}_0T^{-1}=H^{-\nu}_0$, and the operators $H^{+}_0$ and $H^{-}_0$ get interchanged.
On the other hand, if one would (formally!) interpret the Pauli matrices as components
of a genuine spin\cite{PRL2011mecklenburg} (i.e. an angular momentum), the time reversal operator would
read $\tilde T=\sigma_{y}C$ giving $\tilde TH^{\nu}_0\tilde T^{-1}=H^{\nu}_0$.
The parity operator $P$ flips the sign of the spatial coordinates interchanging
the two sublattices and, similar to $T$, fulfills $PH_0^{\nu}P^{-1}=H_0^{-\nu}$.
Thus, the initial Hamiltonian $H^{+}_0+H^{-}_0$ is $PT$-invariant
but, as we shall see, the exchange interactions can break either invariance.
It also is important that the pseudospin is not linked with
the internal magnetic moment of an electron and does not directly interact
with the external magnetic field prohibiting Stern--Gerlach type experiments.
In contrast to that, we predict situations where the pseudospin 
manifests itself in observable quantities and can be detected in transport 
as well as optical measurements on graphene.

First of all we show that the exchange electron-electron interaction can 
alter the pseudospin orientation in a very broad range.
In an eigenstate of $H^\nu_0$ the pseudospin is always in the $xy$-plane.
As we shall see shortly,
the exchange interactions can turn the pseudospin texture
to the out-of-plane phase with the out-of-plane angle
depending on the absolute value of the particle momentum.
This is due to the huge negative contribution to the Hartree--Fock ground 
state energy from the valence band (i. e. ``antiparticle'' states) 
which cannot be neglected in graphene because of the zero gap
(i. e. zero effective mass of carriers) and large effective fine structure 
constant $\alpha^*=e^2/(\varepsilon \hbar v_0)$
where $\varepsilon$ is the dielectric constant depending on the 
environment\cite{PRL2008jang}.
The exchange contribution to the ground state energy has previously been
studied in both monolayer and bilayer graphene
regarding properties such as the electronic 
compressibility\cite{NatPhys2008martin}
and ferromagnetism\cite{PRB2005peres,PRL2007barlas,PRB2006nilsson}, but the 
importance of the interplay
between pseudospin and electron-electron interactions has been recognized 
only in \cite{PRB2008min} where single layer graphene was mentioned in passing.

Having established the possibility to create an out-of-plane pseudospin 
orientation by means of the exchange interaction,
we apply the Boltzmann approach to derive the electrical conductivity tensor 
which turns out to have Hall components even though the external magnetic 
field is absent.
The mechanism of this phenomenon is intimately linked to the 
pseudospin-momentum 
coupling which can be read out immediately from the Hamiltonian $H^\nu_0$.
Similar to the skew scattering of electrons on impurities in 
spin-orbit coupled systems partly responsible for the anomalous Hall 
effect\cite{RMP2010nagaosa,JP2008sinitsyn},
the carriers in graphene do also skew to one side of the Hall bar as long
as their pseudospin has non-zero out-of-plane component.
This effect has been intensively 
studied\cite{PRL2006sinitsyn,PRB2007sinitsyn,arxive2011tse}
assuming that the out-of-plane component occurs due to the band gap opened 
by spin-orbit coupling\cite{PRL2006sinitsyn}
which, however, seems to be weak in graphene\cite{Nature2007geim}.
We emphasize that neither spin-orbit coupling nor an external magnetic field
is necessary to obtain a Hall current in graphene being in the pseudospin 
out-of-plane phase.

Experimental manifestations of the pseudospin are not limited to the 
electron skew scattering phenomenon
but can also be seen in the interband optical absorption.
Performing optical measurements on graphene\cite{SST2010orlita} one can obtain 
direct information regarding
conduction and valence band states without advanced sample processing 
necessary for transport investigations.
Moreover, the peculiar properties discovered so far make graphene a very 
promising material
for optoelectronic applications\cite{Natphot2010review}.
Optical absorption via the direct interband optical transitions in graphene 
has been investigated in \cite{Science2008nair} 
but the mechanism considered there lies essentially in the 
two-dimensional nature and gapless
electronic spectrum and does nor directly involve the pseudospin orientation.
Here we show that, due to the out-of-plane pseudospin orientation, 
the interband absorption can be substantially reduced or enhanced
as compared to its universal value $\pi e^2/ \hbar c$ 
just by switching the helicity of the circularly polarized light.

\begin{figure}
 \includegraphics[width=\columnwidth]{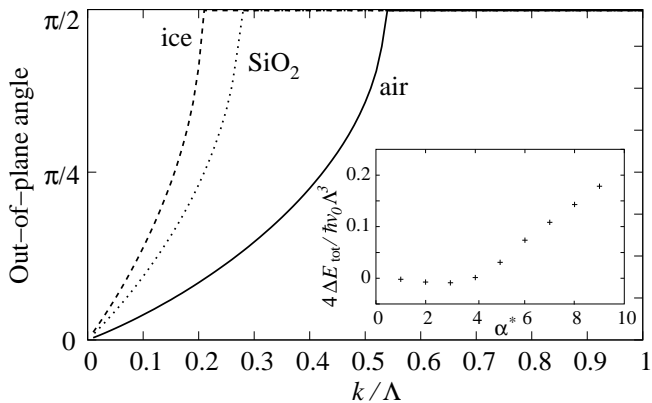}
 \caption{\label{fig1} The pseudospin out-of-plane angle $\vartheta(k)$ 
for different environments numerically calculated from Eq.~(\ref{x}).
The corresponding values of the substrate-dependent effective fine structure 
constant $\alpha^*$ are 
taken from Ref.~\cite{PRL2008jang}.
The inset shows the total ground state energy difference (\ref{Etot}) 
between the in-plane and out-of-plane phases
for different effective fine structure constant 
$\alpha^*=e^2/\varepsilon \hbar v_0$.
Increasing $\alpha^*$ makes the out-of-plane phase more preferable.}
\end{figure}

{\em Exchange interactions.}
The Coulomb exchange Hamiltonian is given by
\begin{equation}
\label{ex}
H^\nu_\mathrm{exch}(\mathbf{k})=-\sum\limits_{\kappa'}\int\frac{d^2k'}{4\pi^2}
U_{|\mathbf{k}-\mathbf{k}'|} |\chi^\nu_{\kappa' k'}\rangle\langle\chi^\nu_{\kappa' k'}|
\end{equation}
with $U_{|\mathbf{k}-\mathbf{k}'|}=2\pi e^2/\varepsilon|\mathbf{k}-\mathbf{k}'|$  
and $\kappa'=\pm$ being the band index with $\kappa=+$ for the conduction band.
The intervalley overlap is assumed to be negligible, and the eigenstates of
$H^\nu=H^\nu_0 +H^\nu_\mathrm{exch}$ can be formulated as
$\Psi_{\mathbf{k}\kappa}^\nu(\mathbf{r})={\mathrm e}^{i\mathbf{kr}}|\chi_{\pm k}^\nu\rangle$
with spinors 
$|\chi_{+ k}^\nu\rangle=(\cos\frac{\vartheta_k}{2}, \nu\sin\frac{\vartheta_k}{2}\mathrm{e}^{\nu i\varphi})^T$,
$|\chi_{- k}^\nu\rangle=(\sin\frac{\vartheta_k}{2}, -\nu\cos\frac{\vartheta_k}{2}\mathrm{e}^{\nu i\varphi})^T$,
and $\tan\varphi=k_y/k_x$.
Thus, a non-zero out-of-plane pseudospin component 
corresponds to $\vartheta_k\neq \pi/2$.
To diagonalize $H^\nu$ the following $\nu$-independent equation for $\vartheta_k$ 
must be satisfied\cite{PRB2008juri}
\begin{eqnarray}
\nonumber &&
\hbar v_0 k\cos\vartheta_k+\sum\limits_{\kappa'}
\int\frac{d^2k'}{8\pi^2}\kappa'U_{|\mathbf{k}-\mathbf{k}'|}
\left[\cos\vartheta_{k'}\sin\vartheta_k- \right. 
 \\
&&\left. -\sin\vartheta_{k'}\cos\vartheta_k\cos(\varphi'-\varphi) \right] =0,
\label{thetak}
\end{eqnarray}
where the integration goes over the occupied states.
Note that the conduction and valence states are entangled,
and the latter cannot be disregarded even at positive Fermi energies. Thus,
in order to evaluate the integrals in Eq.~(\ref{thetak}) a momentum
cut-off $\Lambda$ is necessary. Its value $\simeq 0.1\mathrm{nm}^{-1}$ 
is usually chosen to keep the number of states
in the Brillouin zone fixed\cite{PRB2005peres}, but our outcomes do not 
depend on any particular choice of $\Lambda$.
Substituting $x=k/\Lambda$ we arrive at
\begin{eqnarray}
\label{x} &&
\frac{4\pi x\cos\vartheta_k}{\alpha^*}  = \\
\nonumber &&   \int\limits_0^{2\pi} d\varphi'
\int\limits_{k_F/\Lambda}^{1}dx'x'
\frac{\cos\vartheta_{k'}\sin\vartheta_k 
- \sin\vartheta_{k'}\cos\vartheta_k\cos\varphi'}
{\sqrt{x^2+x'^2-2xx'\cos\varphi'}}.
\end{eqnarray}
The momentum cuf-off is obviously much larger than the Fermi momentum $k_F$
at any reasonable electron doping, and therefore we can set the lower 
integral limit to zero.
Besides a trivial solution with $\vartheta_0=\pi/2$ independent of $k$, there are
non-trivial ones $\vartheta_1=\vartheta(k)$ and $\vartheta_2=\pi-\vartheta(k)$
with $\vartheta(k)$ shown in Fig.~\ref{fig1} for different $\alpha^*$.
The solutions $\vartheta_0$ and $\vartheta_{1,2}$ represent to two phases with different total ground 
state energies $E_\mathrm{tot}^\mathrm{in}$ ($E_\mathrm{tot}^\mathrm{out}$) for the in-plane 
(out-of-plane) pseudospin phase. The difference 
$\Delta E_\mathrm{tot} =E_\mathrm{tot}^\mathrm{in}-E_\mathrm{tot}^\mathrm{out}$
per volume for a given spin and valley reads
\begin{widetext}
\begin{equation}
\label{Etot}
\frac{\Delta E_\mathrm{tot}}{\hbar v_0 \Lambda^3} 
= -\int\limits_0^1 \frac{dx'}{2\pi} x'^2(1-\sin\vartheta_{k'})
-\alpha^*\int\limits_0^{2\pi} d\varphi \int\limits_0^{2\pi} d\varphi'
\int\limits_0^{1}dx\int\limits_0^{1}dx'xx'
\frac{(1-\sin\vartheta_{k'}\sin\vartheta_k)
\cos(\varphi'-\varphi)-\cos\vartheta_{k'}\cos\vartheta_k}
{32\pi^3\sqrt{x^2+x'^2-2xx'\cos(\varphi-\varphi')}}.
\end{equation}
\end{widetext}
The energy difference  for $\alpha^*\sim 1$ is small because the integrand in Eq.~(\ref{Etot})
is always multiplied by $x'$ and therefore vanishes at $x'\to 0$, but at larger $x'$ the 
$\vartheta_{k'}$ gets close to $\pi/2$,
and the integrand vanishes again.
The Inset in Fig.~\ref{fig1} shows, however, that strong electron-electron 
interactions make the out-of-plane
phase energetically preferable.
The estimates of $\alpha^*$ for clean graphene vary from $2$ 
(Ref.~\cite{PRL2008jang}) to $2.8$ (Ref.~\cite{PRB2005peres})
and are on the borderline of the out-of-plane phase.
Moreover, the presence of disorder can change this qualitative picture 
essentially\cite{PRB2005peres}.
Most importantly, Eq.~(\ref{Etot}) is valid for {\em both} valleys and {\em both} solutions $\vartheta_{1,2}$.
Thus, it is possible to choose either the same or opposite solutions for
two valleys. The former choice breaks the parity invariance whereas
the latter one does so with the time reversal symmetry.
Both cases are worthy of consideration.

\begin{figure}
 \includegraphics[width=\columnwidth]{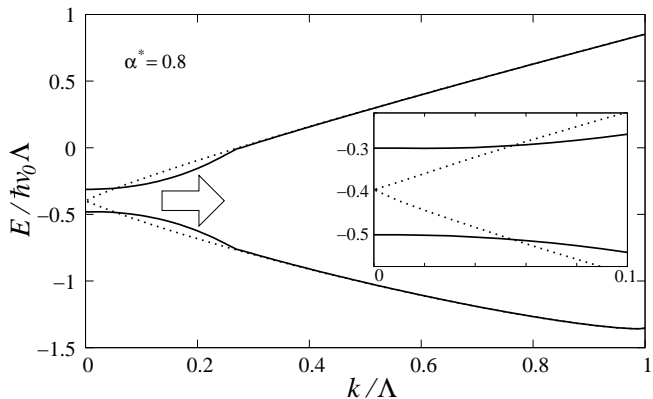}
 \caption{\label{fig2} The dispersion law $E_\kappa(k)$ in the in-plane 
(dashed curves) and out-of-plane (solid curves) phases for  
$\alpha^*=0.8$ corresponding to $\mathrm{SiO_2}$ substrate\cite{PRL2008jang}.
The curves for both phases coincide for momenta larger than a certain 
critical value where $\vartheta_k=\pi/2$
becomes independent of $k$, see Fig.~\ref{fig1}. The inset shows the gap 
region in detail.}
\end{figure}

The single-particle spectrum is independent of the valley index and given by
\begin{eqnarray}
\nonumber&& 
\frac{E_{\kappa }(x)}{\hbar v_0 \Lambda}=\kappa x\sin\vartheta_k 
- \frac{\alpha^*}{4\pi} \int\limits_0^{2\pi} d\varphi' \int\limits_0^{1}dx'x'\\
&& \label{E}
\times \frac{1-\kappa(\cos\vartheta_{k'}\cos\vartheta_k 
+ \sin\vartheta_{k'}\sin\vartheta_k\cos\varphi')}
{\sqrt{x^2+x'^2-2xx'\cos\varphi'}},
\end{eqnarray}
and the group velocity can be written as $\mathbf{v}_\kappa
=v_{\kappa }(\cos\varphi,\sin\varphi)^T$ with $v_\kappa$ being
\begin{eqnarray}
\nonumber&& 
\frac{v_{\kappa }}{v_0}=\kappa\sin\vartheta_k + \frac{\alpha^*}{4\pi} \int\limits_0^{2\pi} d\varphi' \int\limits_0^{1}dx'x'x (1-\cos\varphi')\\
&& \label{v} 
\times \frac{1-\kappa(\cos\vartheta_{k'}\cos\vartheta_k + \sin\vartheta_{k'}\sin\vartheta_k\cos\varphi')}
{(x^2+x'^2-2xx'\cos\varphi')^{\frac{3}{2}}}.
\end{eqnarray}
The dispersion law (\ref{E}) is depicted in Fig.~\ref{fig2} for graphene 
placed on $\mathrm{SiO_2}$ substrate.
The interactions shift the bands down to lower energies
and change the density of states
but, most importantly, they open a gap  \cite{PRL2001khveshchenko} between the valence and 
conduction band
as soon as the system changes to the pseudospin out-of-plane one phase.
The gap at $k=0$ equals 
$\frac{e^2\Lambda}{\varepsilon}\int\limits_0^{1}dx'\cos\vartheta_{k'}$.
Note that the group velocity (\ref{v}) vanishes at small momentum 
$k/\Lambda\ll 1$
as long as the system is in the out-of-plane phase
corresponding to the almost flat bands  close to $k=0$ shown in the inset of 
Fig.~\ref{fig2}.
From now on we assume n-doping so that the Fermi energy is always higher than
the bottom of the conduction band.


{\em Zero-field Hall current.}
To describe the Hall conductivity due to skew scattering we utilize 
the semiclassical Boltzmann
approach which allows a physically transparent interpretation of this
mechanism\cite{JP2008sinitsyn,PRB2007sinitsyn}.
In general the anomalous Hall conductivity contributions can be classified 
by their mechanism:
(i) The intrinsic contribution is due to the anomalous velocity of carriers
(being non-diagonal with respect to the band index \cite{EPL2008trushin})
which is coupled to the equilibrium part of the distribution function.
(ii) The side-jump contribution follows from coordinate shifts during 
scattering events.
It occurs in the non-equilibrium part of the distribution function as 
well as in the anomalous velocity\cite{JP2008sinitsyn,PRB2007sinitsyn}.
(iii) The skew scattering contribution is independent of the coordinate 
shift and of the anomalous velocity.
It occurs when the scattering rate is asymmetric with respect to the 
initial and final states
and, therefore, must be considered beyond the first Born approximation
The first two conductivity contributions do not depend on disorder but
on the out-of-plane angle $\vartheta_k$ and can be adopted from 
\cite{PRB2007sinitsyn}. Here, we focus on the skew scattering contribution
which can be described using the interband incoherent 
Boltzmann equation where the anomalous velocity is neglected but the scattering probability
is calculated up to the third order in the short-range scattering 
potential with the momentum-independent Fourier transform $V$.
In linear order in the homogeneous electric field  ${\mathbf E}$ this 
equation reads
$-e{\mathbf E} \mathbf{v}_k \left[-\partial f^0(E_{k})/ \partial
E_{k}\right]=I[f^1_k]$, where
$f^0(E_{k})$ is the Fermi-Dirac function,
$f^1_\mathbf{k}$ is the non-equilibrium addition,
and $\mathbf{v}_k$, $E_{k}$ are given by Eqs.~(\ref{v},\ref{E}) with $\kappa=+$.
The collision integral can be written as
$I[f^1_\mathbf{k}]=\int\frac{d^2k'}{(2\pi)^2}
w_{\mathbf{k}\mathbf{k'}}(f^1_\mathbf{k'}-f^1_\mathbf{k})$
with $w_{\mathbf{k}\mathbf{k'}}$ being the scattering probability.
We divide $w_{\mathbf{k}\mathbf{k'}}$ into two parts.
The first one is proportional to the cosine of the scattering angle
and calculated up to the second order in $V$.
The second one is proportional to the sine of the scattering angle
and calculated up to the third order in $V$.
These two parts correspond to the conventional and skew scattering respectively
which can be alternatively expressed in terms of the momentum relaxation times,
cf. Ref.\cite{PRL2006sinitsyn}
\begin{eqnarray}
\nonumber &&
(\tau_\parallel^\nu)^{-1}=n_ikV^2(1+3\cos^2\vartheta_k^\nu)/(4\hbar^2v_k),\\
&& 
(\tau_\perp^\nu)^{-1}=\nu n_ik^2V^3\cos\vartheta_k^\nu\sin^2\vartheta_k^\nu/(8\hbar^3v_k^2).
\label{tau1}
\end{eqnarray}
Here, $n_i$ is the concentration of such scatterers.
Since $\tau_\perp^\nu \propto 1/V^3$ whereas $\tau_\parallel^\nu \propto 1/V^2$ it is 
natural to assume $\tau_\perp^\nu \gg \tau_\parallel^\nu$,
and the Hall conductivity for a given valley can be estimated as 
$\sigma_{yx}^\nu\approx\sigma_{xx}\tau_\parallel^\nu/\tau_\perp^\nu\mid_{k=k_F}$ which can vary in a 
quite broad but finite range because neither of $\tau$'s diverges at low doping thanks to the $k$-dependent group velocity (\ref{v}).
Note that $K$ and $K'$ contribute identically to the total Hall
conductivity $\sigma_{yx}=\sum_{\nu}\sigma_{yx}^\nu$ if the out-of-plane 
pseudospin polarization is opposite in the two valleys,
i. e. $\vartheta_1$ and $\vartheta_2$ are assigned to $\vartheta_k^+$ and $\vartheta_k^-$ respectively,
and the time reversal invariance is broken by the exchange interactions.
On the other hand, if the out-of-plane  pseudospin polarization is the same in both valleys
(i. e. either of $\vartheta_{1,2}$ is assigned to both $\vartheta_k^\pm$ breaking
the parity invariance) the Hall currents in the two valleys have opposite directions
resulting in the {\em valley} Hall effect\cite{PRL2007xiao} ---
another analog of the well known spin Hall effect\cite{PRL2006sinitsyn}.

{\em Interband optical absorption.}
From $H^\nu_0$ one can deduce 
the following interaction Hamiltonian between the electromagnetic wave 
and carriers in graphene
$H^\nu_\mathrm{int}= \frac{e v_0}{c}(\nu\sigma_x A_x+\sigma_y A_y)$
which couples the vector potential $\mathbf{A}$ 
and pseudospin $\vec{\sigma}$. As consequence,
the inter-band transition matrix elements turn out to be sensitive to
the light polarization and pseudospin orientations in the initial and 
final states.
To be specific we assume monochromatic light of frequency $\omega$, normal 
incidence (i.e. zero momentum transfer from photons to electrons), and 
circular polarization (fulfilling $A_x=\pm iA/\sqrt{2}$, $A_y=A/\sqrt{2}$).
The probability to excite an electron from the valence band to an 
unoccupied state in the conduction band
can be calculated using the golden-rule.
Finally, the absorption $P^\nu$ can be calculated as a ratio between 
the total electromagnetic power $W_a$ absorbed
by graphene per unit square and the incident energy flux 
$W_i=\omega^2 A^2/4\pi c$.
Then, the optical absorption for $K$ valley ($\nu=+$) reads 
\begin{equation}
\label{P}
P^+=\frac{\pi e^2}{\hbar c}\frac{4\Lambda v_0}{\omega} \int\limits_0^\infty dx x
\left\{ \begin{array}{c}
\sin^4\frac{\vartheta_k^+}{2} \\ 
\cos^4\frac{\vartheta_k^+}{2}
\end{array} \right\} \delta\left(\frac{E_+ - E_- 
- \hbar\omega}{\hbar v_0 \Lambda}\right),
\end{equation}
where the multipliers $\sin^4(\vartheta_k^+/2)$ and $\cos^4(\vartheta_k^+/2)$ are
for two opposite helicities of light, and for K'-valley they are interchanged.
If the out-of-plane pseudospin polarization is chosen to be opposite in the two valleys,
then the total absorption $P=\sum_{\nu} P^\nu$ at small $k/\Lambda$ turns
out to be sensitive to the helicity of light:
It is substantially reduced for one and facilitated for another.
Moreover changing the excitation energies $\hbar\omega$ we can investigate 
the dependence $\vartheta(k)$ shown in Fig.~\ref{fig1}.
If the out-of-plane pseudospin polarization is chosen to be the same in both valleys,
then the total absorption does not depend on the radiation helicity but
the two valleys turn out to be differently occupied by the photoexcited carriers
which is interesting effect on its own\cite{PRL2007xiao}.
In the in-plane phase with $\vartheta=\pi/2$ the total absorption does not depend on
light polarization, and in the non-interacting limit it equals to the universal 
value $\frac{\pi e^2}{\hbar c}$, as expected\cite{Science2008nair}.

{\em Conclusions.}
We have demonstrated that the pseudospin being until now rather
uncontrollable and almost unmeasurable quantity can be
``unfrozen'' by the exchange electron-electron interactions (\ref{ex})
and play an essential role in optical and transport properties of graphene.
We hasten to say that the Hartree-Fock approximation employed here
has generically a tendency to overestimate ordering such as the
pseudospin out-of-plane polarization. 
We believe, however, that the pseudospin eigenstates $|\chi_{\kappa k}^\nu\rangle$
derived above are much more robust because their special pseudospin-momentum
entangled structure stems from the free Hamiltonian $H^\nu_0$,
and the electron-electron interactions do only modify it
making our predictions reliable at the qualitative level. 
From this point of view the pseudospin can be seen as an additional degree of freedom
similar to the true spin but unaffected by the magnetic field directly.
Having this similarity in mind one can think about pseudospin
ferromagnetism\cite{PRB2008min}, pseudospin accumulation at the sample's edge
by means of the zero-field Hall current,
pseudospin selectivity in the optical absorption (\ref{P}),
and, probably, pseudospin filtering and switching.
In a more distant future one can imagine some
useful effects based on the pseudospin polarization like
an all-electrical counterpart for GMR which
is obviously very promising for application.
This Letter should be seen as a first step in this direction.

This work was supported by DFG via GRK 1570.

\bibliography{graphene.bib,optical.bib}

\end{document}